\documentclass[runningheads]{llncs}
\usepackage[T1]{fontenc}
\usepackage{graphicx,verbatim}
\usepackage{hyperref}
\usepackage{amsmath}
\usepackage{amssymb}
\usepackage{multirow}
\usepackage{tabularx}
\usepackage{color}

\begin{document}
\title{FedWSIDD: Federated Whole Slide Image Classification via Dataset Distillation}
%
\author{Haolong Jin\inst{1} \and
Shenglin Liu\inst{1} \and Cong Cong\inst{2} \and Qingmin Feng\inst{1} \and Yongzhi Liu\inst{1} \and Lina Huang\inst{1} \and Yingzi Hu\inst{1}}
\authorrunning{Jin et al.}
\institute{Union Hospital, Tongji Medical College, Huazhong University of Science and Technology \and
Centre for Health Informatics, Australian Institute of Health Innovation, Macquarie University, Sydney, NSW 2113, Australia\\
\email{jhl\_bme@hust.edu.cn}\\}

\maketitle              
\begin{abstract}
Federated learning (FL) has emerged as a promising approach for collaborative medical image analysis, enabling multiple institutions to build robust predictive models while preserving sensitive patient data.
In the context of Whole Slide Image (WSI) classification, FL faces significant challenges, including heterogeneous computational resources across participating medical institutes and privacy concerns. To address these challenges, we propose FedWSIDD, a novel FL paradigm that leverages dataset distillation (DD) to learn and transmit synthetic slides.
On the server side, FedWSIDD aggregates synthetic slides from participating centres and distributes them across all centres. On the client side, we introduce a novel DD algorithm tailored to histopathology datasets which incorporates stain normalisation into the distillation process to generate a compact set of highly informative synthetic slides. 
These synthetic slides, rather than model parameters, are transmitted to the server. After communication, the received synthetic slides are combined with original slides for local tasks.
Extensive experiments on multiple WSI classification tasks, including CAMELYON16 and CAMELYON17, demonstrate that FedWSIDD offers flexibility for heterogeneous local models, enhances local WSI classification performance, and preserves patient privacy.
This makes it a highly effective solution for complex WSI classification tasks. The code is available at \href{https://github.com/f1oNae/FedWSIDD}{FedWSIDD}.

\keywords{Federated Learning \and Dataset Distillation \and WSI classification.}

\end{abstract}
\section{Introduction}
Computational pathology aims to enhance the precision of histopathological tissue examination and it is crucial for accurate disease diagnosis. Deep learning has demonstrated significant success in various digital pathology tasks; however, Whole Slide Images (WSIs) pose unique challenges. Their enormous size makes direct model training impractical, and the labour-intensive nature of annotation often limits available labels to the slide level.  
To address these issues, weakly-supervised Multiple Instance Learning (MIL) frameworks \cite{campanella2019clinical,chen2022scaling} are commonly used, leveraging slide-level ground truth labels while treating each WSI as a collection of patches. 

As with many machine learning applications, incorporating diverse data that reflect variations in patient populations and data collection protocols can significantly enhance MIL performance \cite{lu2022federated}. However, sharing medical data poses challenges, including regulatory and legal constraints, as well as technical obstacles like the high costs of data transfer and storage. Federated learning (FL) provides feasible solutions to the above issues by enabling algorithms to be trained on decentralised data across multiple institutions, allowing each institution to retain its data securely. 
FL-based WSI classification methods have been actively developed in recent years \cite{huang2024federated}. Though effective, they focus on patch-level classification which requires precise patch-level annotations. However, such annotations are usually hard to obtain in real clinical scenarios. Thus, FedHisto \cite{lu2022federated} was proposed to tackle the more practical slide-level classification task where the MIL classifiers are shared between the centres and server. Huang \emph{et al.}~\cite{huang2024federated} further enhanced the local model performance by introducing a dynamic parameter updating mechanism. However, these methods assume that centres have similar computational resources and identical MIL architectures, limiting their application in settings with heterogeneous MIL models.

Accordingly, we propose FedWSIDD, a novel federated learning (FL) algorithm for weakly supervised WSI classification that incorporates dataset distillation. 
To ensure compatibility across different MIL models, rather than employing model augmentation~\cite{zhang2023accelerating} or maintaining a diverse MIL model pool~\cite{cazenavette2022dataset}, we exploit the fact that pre-extracted features are MIL-agnostic. Hence, we explicitly design a dataset distillation \cite{wang2018dataset,cong2024dataset} algorithm that matches the feature space between real and synthetic patches, enabling a more adaptable and efficient dataset distillation process.
Furthermore, to mitigate the training challenges caused by the inherent heterogeneity in the appearance of H\&E-stained tissue slides, we integrate stain normalisation \cite{cong2022colour,cong2021texture,macenko2009method} into the dataset distillation process.
At each participating medical centre, dataset distillation is performed to generate synthetic slides, which are then transmitted to the central server. The server aggregates and redistributes these synthetic slides to all centres. Each centre subsequently incorporates the updated synthetic slides from other centres as local data augmentation in the next training round.
Our contributions can be summarised as follows:
\begin{itemize}
    \item We introduce FedWSIDD, a novel federated learning paradigm for weakly supervised WSI classification, where each centre learns and shares synthetic WSIs, enabling flexible MIL selection based on local computational resources.
    \item We design a dataset distillation algorithm tailored for WSI datasets, which integrates stain normalisation into the learning process of synthetic patches, enhancing the quality of the synthetic slides and leading to enhanced downstream classification performance.
    \item Extensive experiments demonstrate the effectiveness of FedWSIDD. Compared to state-of-the-art FL methods, FedWSIDD achieves superior performance across various settings, including both homogeneous and heterogeneous local models.
\end{itemize}

\section{Method}
\subsection{Preliminary}
\begin{figure*}[t]
    \centering
    \includegraphics[width=0.9\textwidth]{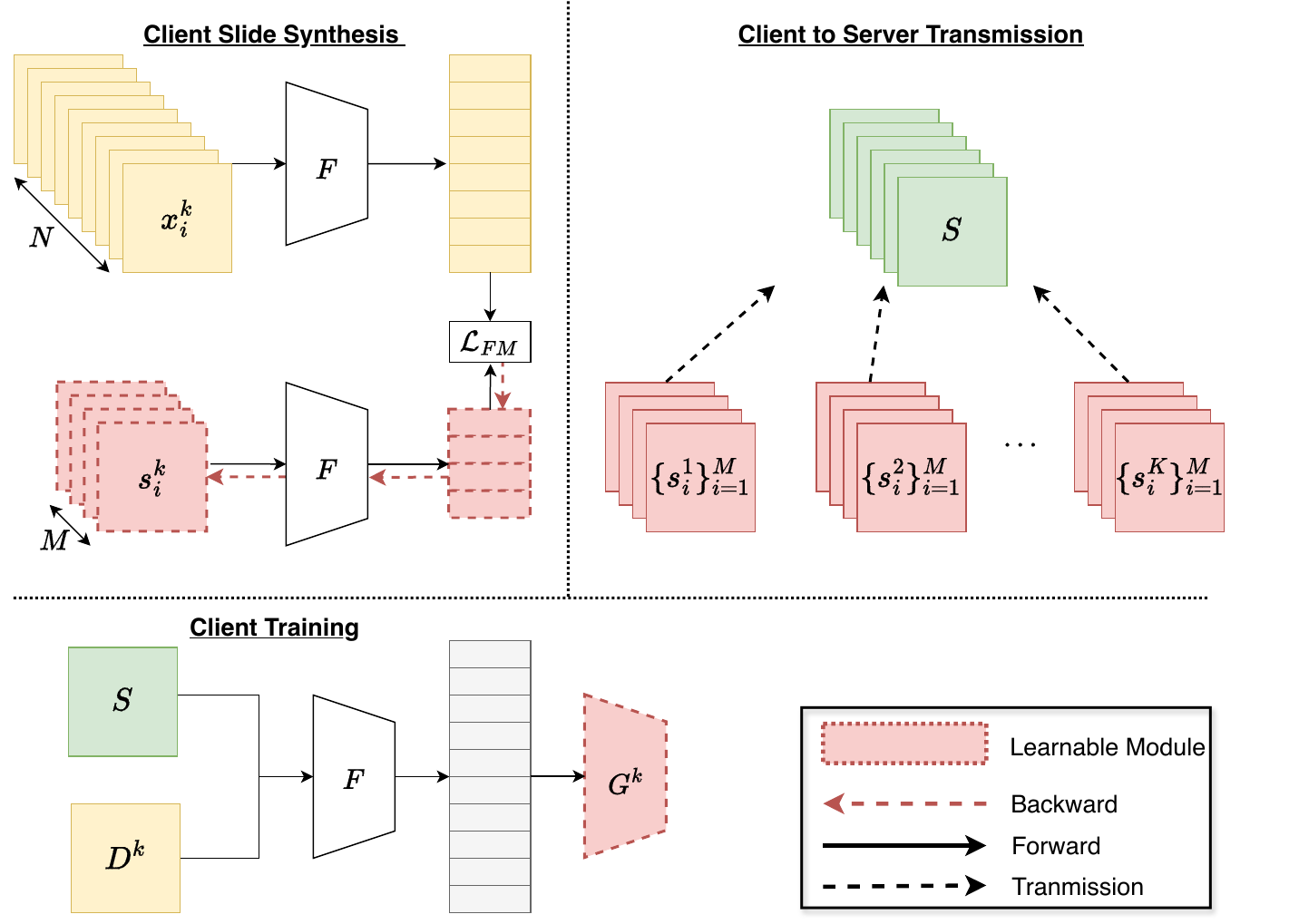}
    \caption{FedWSIDD pipeline. Each client first generates a set of synthetic slides which are then transmitted to the server. The server gathers synthetic slides from all participating clients and distributes the aggregated synthetic slides to them. Upon receiving the synthetic slides, clients merge them with their original real slides for local training.}
    \label{fig:framework}
\end{figure*}
Given a WSI dataset consisting of $N$ WSI slides $\{x_{i}, y_{i}\}_{i}^{N}$, we first crop $x_{i}$ into a set of patches $\{c_{t}^{i}\}_{t=1}^{T}$, where $T$ could vary across WSIs. Then a pretrained feature extractor ($\mathcal{F}$) is used to convert $c_{t}^{i}$ to patch embedding $e_{t}^{i}=\mathcal{F}(c_{t}^{i})$. This is usually done in an offline manner where $e_{t}^{i}$ is then aggregated to form a bag representation that is handled by the MIL classifier (${G}$) to obtain the slide-level prediction.

The standard Federated WSI classification involves multiple medical centres working collaboratively to enhance the performance of a global MIL model. In practical scenarios, different medical centres may have different computational resources and wish to adopt different feature extractors or MIL classifiers. Thus, we focus on the personalised federated WSI classification to tackle this heterogeneous local model scenario. Supposing we have $K$ centres where each consists of their own WSI datasets ($\{D^1,D^2,\cdots,D^K\}$) and the task can be formulated as:
\begin{equation}
    \mathop{\arg\min}_{\{{G}^{k}\}_{k=1}^{K}} \mathcal{L}=\sum_{k=1}^{K}\frac{|D^k|}{|D|}\mathbb{E}_{(x,y)\in D^k}\mathcal{L}({G}^{k}; x, y)
    \label{eq:mk}
\end{equation}
here $D$ represents the combined dataset from all centres, ${G}^{k}$ denotes the MIL classifier selected by centre $k$ and $\mathcal{L}$ is a task-specific loss function, and different MIL methods define $\mathcal{L}$ in various forms. For example, in CLAM \cite{lu2021data}, $\mathcal{L}$ comprises both slide-level and patch-level loss components.
If all centres choose to use the same MIL classifier, this scenario becomes a homogeneous local model setup and standard FL methods can be applied.

\subsection{Federated WSI Learning with Synthetic Slides}
We illustrate the pipeline of FedWSIDD in Fig.~\ref{fig:framework}. Since each centre may have different computational resources and employ a distinct MIL classifier $G^k$, transmitting model weights or gradients, as done in traditional FL methods~\cite{lu2022federated}, is neither practical nor efficient. To overcome this limitation, we propose generating a set of synthetic slides and transmitting $\{s_{i}\}^{M}_{i=1}$, where $M$ represents the number of synthetic slides per class and satisfies $M \ll N$, instead of exchanging model weights.

Specifically, each participating centre $k$ first performs local synthetic slide generation using dataset distillation (see Sec.~\ref{sec:dd} for details) to produce $M$ synthetic slides, denoted as $\{s^{k}_{i}\}^{M}_{i=1}$. These synthetic slides are then uploaded to the server, where the global synthetic slides ($S$) are aggregated as:
\begin{equation}
    S=\mathop{\cup}_{k\in K}\{s_{i}^{k}\}_{i=1}^{M}
    \label{eq:tildeP}
\end{equation}
The global synthetic slides $S$ are subsequently distributed back to the participating centres, where they are merged with real local slides for local MIL training.

\subsection{Local Synthetic Slides with Dataset Distillation}
\label{sec:dd}
Generating synthetic WSIs directly from real slides is highly challenging due to their gigapixel resolution. To address this, we represent each synthetic slide \( s_{i} \) as a collection of synthetic patches, \emph{i.e.,} \( s_{i}=\{p_j\}_{j=1}^{B} \), where \( p_j \in \mathbb{R}^{C\times H\times W} \) and \( B \ll T \), with each patch treated as a set of learnable parameters.  
To ensure compatibility across different MIL models, we leverage the fact that pre-extracted features are ML-agnostic. Thus, we explicitly match the feature space between real and synthetic patches, enabling a more adaptable and efficient dataset distillation process.
Specifically, during each distillation iteration, we iterate through all classes. For each class, we randomly sample one real slide ($x_j$) and one synthetic slide ($s_j$).
To mitigate training difficulties caused by the heterogeneity of H\&E-stained tissue slides, we incorporate stain normalisation~\cite{macenko2009method} into the training process. We apply stain normalisation \cite{cong2024dataset} to both the real patches $\{c_{t}^{j}\}_{t=1}^{T}$ within $x_j$ and the synthetic patches $\{p_{b}^{j}\}_{b=1}^{B}$ within $s_j$ and perform feature matching using:
\begin{equation}
    \mathop{\arg\min}_{\{p_b^j\}_{b=1}^{B}} \mathcal{L}_{FM} = \left[ \left\| \frac{1}{T}\sum_{i=0}^T\mathcal{F}(c_t^j) - \frac{1}{B}\sum_{b=0}^{B}\mathcal{F}(p_b^j) \right\|^2 \right],
\end{equation}
Note that, unlike previous works~\cite{huang2024overcoming} that apply dataset distillation to federated learning, we are the first to successfully extend this concept to WSI classification, providing a potential solution for real-world clinical applications. Furthermore, we incorporate stain normalisation into the learning process of synthetic patches, enhancing dataset distillation algorithms specifically for histopathology images. Empirically, we demonstrate that FedWSIDD outperforms these existing approaches on WSI datasets.

\section{Experiments}
\subsection{Datasets}
\noindent \textbf{CAMELYON16} \cite{bejnordi2017diagnostic} is for detecting breast cancer metastasis in lymph node sections. The task is binary classification, distinguishing between normal and tumour slides. The dataset includes 399 WSIs from two medical centres: RUMC (C1), with 169 slides for training (99 normal/70 tumour) and 74 slides for testing (50 normal/24 tumour); and UMCU (C2), contributing 101 slides for training (60 normal/41 tumour) and 55 slides for testing (31 normal/24 tumour).

\noindent \textbf{CAMELYON17} \cite{bandi2019detection} consists WSIs collected from five medical centres. As test labels are not officially provided, we used the training set (500 slides from 100 patients) for our experiments. The task is to classify each slide into one of four categories: negative (neg), isolated tumour cells (itc), micro-metastases (micro), and macro-metastases (macro). Each centre (C1–C5) includes 20 patients, with five slides per patient. The class distribution across centres is as follows: C1: 11 itc, 10 micro, 15 macro, and 64 neg; C2: 7 itc, 23 micro, 12 macro, and 58 neg; C3: 2 itc, 8 micro, 15 macro, and 75 neg; C4: 8 itc, 5 micro, 26 macro, and 61 neg; C5: 8 itc, 26 micro, 5 macro, and 61 neg. Within each centre, We randomly allocated 80\% of data for training and 20\% for testing, ensuring that slides from the same patient were assigned to the same split.

\subsection{Implementation Details}
We loaded WSIs at 20x for CAMELYON17, and 40x for CAMELYON16. Moreover, we extracted 256x256 patches from CAMELYON16 and CAMELYON17.
For patch feature extraction, we employed a ResNet50 \cite{he2016deep} which is pre-trained on ImageNet \cite{deng2009imagenet}.
Each patch was resized to 224×224 pixels for feature extraction.
For the FL setup, we follow \cite{huang2024overcoming,song2023federated} and conduct a one-shot communication, 1000 distillation rounds and 50 local MIL training rounds. Synthetic slides and MIL models were trained using the Adam optimiser with a learning rate of 0.0003, on a single Nvidia V100 GPU. The synthetic slides were randomly initialised and we set $M=10$ and $B=100$ with a size of 64x64.
For evaluation, we measured test-set accuracy for each centre and the globally averaged accuracy, following the setup in \cite{tan2022fedproto,huang2024overcoming}.
Each experiment was repeated five times with fixed dataset splits to evaluate the impact of random initialisation and training variations, reporting the mean and standard deviation of classification accuracy. To further validate the improvement's significance, we conducted a paired t-test under the null hypothesis that FedWSIDD exhibits no significant performance difference compared to the baselines.


\begin{table*}[t]
\caption{Homogeneous model performance comparison between FedWSIDD and \textit{standard FL} methods on CAMELYON16 (\textbf{CA16}), CAMELYON17 (\textbf{CA17}).}
\centering
\scriptsize
\begin{tabular}{p{1.0cm}p{1.15cm}p{1.22cm}p{1.22cm}p{1.22cm}p{1.22cm}p{1.22cm}p{1.22cm}p{1.22cm}}
\hline
\multicolumn{2}{l}{\textbf{Methods}}    & \multicolumn{1}{c}{Local}        & \multicolumn{1}{c}{FedHisto}                      & \multicolumn{1}{c}{FedDyn}                & \multicolumn{1}{c}{MOON}   & \multicolumn{1}{c}{FedImpro} & \multicolumn{1}{c}{FedMut}  & \multicolumn{1}{c}{FedWSIDD}  \\ \hline
\multirow{4}{*}{\textbf{CA16}} & C1    & $85.2_{\pm 1.0}$  &   ${92.8}_{\pm 1.7}$ &  $92.3_{\pm0.6}$ &$91.0_{\pm1.7}$&$92.1_{\pm1.4}$ &$93.6_{\pm0.5}$ &$\mathbf{94.2}_{\pm0.2}$ \\
                                & C2    & $74.0_{\pm1.6}$ &  $74.8_{\pm1.8}$ &  $84.3_{\pm1.8}$ & $78.0_{\pm1.7}$&$80.3_{\pm0.8}$ &$82.4_{\pm1.2}$ &${84.5}_{\pm0.1}$\\
                                & Avg   & $79.6_{\pm0.6}$ &  $83.8_{\pm 1.6}$  &  $88.3_{\pm0.7}$ & $84.6_{\pm0.3}$&$86.5_{\pm0.8}$&$88.6_{\pm0.6}$ &$\mathbf{90.1}_{\pm0.2}$\\ \hline
\multirow{1}{*}{} &p-value &0.00001&0.0001 &0.002 &0.0002 &0.0008 &0.004 & -\\ \hline
\multirow{7}{*}{\textbf{CA17}} & C1    & ${80.0}_{\pm0.0}$ & ${85.0}_{\pm4.1}$&${80.0}_{\pm0.0}$ &$\mathbf{90.0}_{\pm0.0}$ &${82.8}_{\pm1.5}$&$82.0_{\pm2.0}$  &${86.0}_{\pm2.0}$\\
                                & C2    & ${60.0}_{\pm4.1}$&${71.7}_{\pm2.4}$&${71.7}_{\pm2.4}$&${61.7}_{\pm2.4}$&${70.2}_{\pm4.8}$&$73.0_{\pm2.4}$&$\mathbf{76.0}_{\pm3.7}$ \\
                                & C3    &${86.7}_{\pm2.4}$ &${81.7}_{\pm2.4}$& ${83.3}_{\pm4.7}$&${83.3}_{\pm2.4}$&${88.8}_{\pm0.9}$&$88.0_{\pm2.4}$&$\mathbf{89.0}_{\pm2.0}$ \\
                                & C4    &${60.0}_{\pm0.0}$ &${61.7}_{\pm2.4}$&${58.3}_{\pm2.4}$&${63.3}_{\pm2.4}$&${66.2}_{\pm2.9}$&$70.6_{\pm3.4}$&$\mathbf{71.0}_{\pm2.0}$\\
                                & C5    & ${80.0}_{\pm0.0}$&${81.7}_{\pm4.7}$&${78.3}_{\pm2.4}$&${81.7}_{\pm2.4}$&${83.0}_{\pm1.4}$&$80.6_{\pm3.1}$&$\mathbf{84.0}_{\pm2.0}$\\
                                & Avg   & $73.3_{\pm0.9}$  &$76.3_{\pm0.5}$ &$73.3_{\pm0.9}$&${78.3}_{\pm1.3}$&${79.3}_{\pm1.0}$ &$78.7_{\pm2.4}$& $\mathbf{81.2}_{\pm1.2}$ \\\hline
\multirow{1}{*}{} &p-value &0.0005&0.002&0.0003&0.02&0.03&0.04&- \\ \hline
\end{tabular}
\label{tab:standard}
\end{table*}
\begin{table*}[t]
\caption{Homogeneous model performance comparison between FedDFP and \textit{personalised FL} methods on CAMELYON16 (\textbf{CA16}), CAMELYON17 (\textbf{CA17}).}
\centering 
\scriptsize
\begin{tabular}{p{1.0cm}p{1.15cm}p{1.22cm}p{1.22cm}p{1.22cm}p{1.22cm}p{1.22cm}p{1.22cm}p{1.22cm}}
\hline
\multicolumn{2}{l}{\textbf{Methods}}    & \multicolumn{1}{c}{FedDM}          & \multicolumn{1}{c}{FedProto}            & \multicolumn{1}{c}{DESA}                             & \multicolumn{1}{c}{FedD3} & \multicolumn{1}{c}{FedDGM} & \multicolumn{1}{c}{SGPT}& \multicolumn{1}{c}{{FedWSIDD}}                       \\ \hline
\multirow{4}{*}{\textbf{CA16}} & C1    &$91.9_{\pm0.8}$&${92.8}_{\pm0.6}$  &$92.3_{\pm0.6}$&$93.2_{\pm1.3}$&$92.9_{\pm1.1}$&$90.1_{\pm0.9}$&$\mathbf{94.2}_{\pm0.2}$ \\
                                & C2    &$81.3_{\pm0.6}$& $83.6_{\pm0.9}$ & $83.6_{\pm0.9}$&$82.3_{\pm1.1}$&$83.8_{\pm1.1}$&$83.9_{\pm0.8}$&$\mathbf{84.5}_{\pm0.1}$\\
                                & Avg   &$86.6_{\pm0.7}$& $88.2_{\pm0.7}$&$88.0_{\pm0.8}$ &$87.8_{\pm1.2}$&$88.4_{\pm0.9}$&$87.0_{\pm0.9}$&$\mathbf{90.1}_{\pm0.2}$\\ \hline
\multirow{1}{*}{} &p-value &0.002&0.02 &0.009 &0.006 &0.02 &0.002 & -\\ \hline
\multirow{7}{*}{\textbf{CA17}} & C1    &$82.0_{\pm2.4}$  &$88.3_{\pm2.4}$ &$\mathbf{90.0}_{\pm0.0}$ &$83.0_{\pm2.4}$&$83.0_{\pm2.5}$&$75.0_{\pm0.0}$&${86.0}_{\pm2.0}$\\
                                & C2    & $67.0_{\pm2.4}$ &$75.0_{\pm4.1}$ &$75.0_{\pm0.0}$&$67.0_{\pm6.0}$ &$73.0_{\pm2.7}$&$68.3_{\pm2.4}$&$\mathbf{76.0}_{\pm3.7}$\\
                                & C3    &$82.0_{\pm2.5}$ &${85.0}_{\pm0.0}$ &$76.6_{\pm4.7}$ &${90.0}_{\pm0.0}$&$84.0_{\pm2.2}$&$88.3_{\pm2.4}$&$\mathbf{90.0}_{\pm0.0}$\\
                                & C4    &$64.0_{\pm3.7}$ &${65.0}_{\pm0.0}$&$61.7_{\pm2.4}$&$65.0_{\pm3.0}$&$71.0_{\pm2.0}$&$\mathbf{72.0}_{\pm4.1}$&${71.0}_{\pm2.0}$\\
                                & C5    &$85.0_{\pm0.0}$  &${80.0}_{\pm0.0}$ &$78.3_{\pm2.4}$ &$81.0_{\pm2.0}$&$84.0_{\pm2.0}$&$75.0_{\pm2.4}$&$\mathbf{84.0}_{\pm2.0}$\\
                                & Avg   &$76.0_{\pm1.6}$ &$78.7_{\pm1.1}$&$76.3_{\pm1.5}$ &$77.2_{\pm1.0}$&$78.9_{\pm1.0}$&$76.7_{\pm2.6}$&$\mathbf{81.2}_{\pm0.8}$\\\hline
\multirow{1}{*}{} &p-value &0.02&0.04 &0.01 &0.001 &0.04 &0.02 & -\\ \hline
\end{tabular}
\label{tab:pfl}
\end{table*}
\begin{table}[]
\caption{Heterogeneous model performance comparison between FedDFP and \textit{personalised FL} methods on CAMELYON16 (\textbf{CA16}), CAMELYON17 (\textbf{CA17}).}
\centering
\scriptsize
\begin{tabular}{p{1.0cm}p{1.15cm}p{1.22cm}p{1.22cm}p{1.22cm}p{1.22cm}p{1.22cm}p{1.22cm}p{1.22cm}}
\hline
\multicolumn{2}{l}{\textbf{Methods}}    & \multicolumn{1}{c}{FedDM}          & \multicolumn{1}{c}{FedHE}            & \multicolumn{1}{c}{DESA}                             & \multicolumn{1}{c}{FedD3} & \multicolumn{1}{c}{FedDGM} & \multicolumn{1}{c}{SGPT}& \multicolumn{1}{c}{{FedWSIDD}}                       \\ \hline
\multirow{3}{*}{\textbf{CA16}} & C1   &$78.3_{\pm0.9}$ &$75.2_{\pm2.2}$&$80.8_{\pm2.6}$  &$80.9_{\pm0.8}$&$82.5_{\pm0.9}$&$80.6_{\pm1.2}$&$\mathbf{86.0}_{\pm0.6}$ \\
                               & C2    &$64.4_{\pm1.1}$ &$66.1_{\pm3.9}$&$67.6_{\pm3.4}$ &$70.9_{\pm0.9}$&$73.0_{\pm0.8}$&$71.1_{\pm1.8}$ &$\mathbf{83.7}_{\pm0.9}$ \\
                               & Avg   &$71.4_{\pm0.9}$ &$70.2_{\pm2.0}$ &$75.0_{\pm2.7}$ &$76.0_{\pm1.1}$&$77.6_{\pm1.1}$&$76.2_{\pm1.6}$ &$\mathbf{84.8}_{\pm0.7}$\\ \hline
\multirow{1}{*}{} &p-value &0.00001&0.00004&0.002&0.002&0.006&0.003&- \\ \hline
\multirow{6}{*}{\textbf{CA17}} & C1  &$72.0_{\pm6.2}$&$75.0_{\pm0.0}$&$72.0_{\pm6.2}$&$69.0_{\pm4.2}$&$74.0_{\pm3.2}$&$75.0_{\pm0.0}$ &$\mathbf{77.0}_{\pm2.1}$  \\
                               & C2  &$73.0_{\pm2.1}$&$61.7_{\pm2.3}$&$70.0_{\pm4.1}$&$73.0_{\pm2.1}$&$72.0_{\pm2.1}$&$68.3_{\pm2.4}$ &$\mathbf{73.0}_{\pm2.1}$ \\
                               & C3  &$84.0_{\pm3.7}$&$80.0_{\pm0.0}$&$\mathbf{92.0}_{\pm2.8}$&$84.0_{\pm4.2}$&$85.0_{\pm4.5}$&$88.3_{\pm2.4}$ &$86.0_{\pm2.1}$ \\
                               & C4  &$65.0_{\pm3.7}$&$73.0_{\pm2.4}$&$67.0_{\pm2.4}$ &$69.0_{\pm4.2}$&$67.0_{\pm2.1}$&$71.6_{\pm6.2}$ &$\mathbf{73.0}_{\pm2.1}$  \\
                               & C5  &$74.0_{\pm3.7}$& $75.0_{\pm4.1}$&$75.0_{\pm4.1}$&$72.0_{\pm2.1}$&$72.0_{\pm2.1}$&$75.0_{\pm4.1}$ &$\mathbf{79.0}_{\pm2.1}$ \\
                               & Avg &$73.6_{\pm3.6}$&$73.0_{\pm0.0}$ &$75.7_{\pm2.1}$&$73.2_{\pm5.8}$&$74.2_{\pm4.1}$&$76.7_{\pm2.0}$ &$\mathbf{77.6}_{\pm1.4}$ \\ \hline
\multirow{1}{*}{} &p-value &0.03&0.01&0.03&0.01&0.04&0.04&- \\ \hline
\end{tabular}
\label{tab:heter}
\end{table}

\subsection{Comparison with State-of-the-arts}
\noindent \textbf{Baselines} We compare FedWSIDD with two categories of federated learning (FL) methods.
\textit{Standard FL} involves training a global model through iterative local training and global aggregation. Specifically, we choose FedHisto \cite{lu2022federated}, an enhanced version of FedAvg \cite{mcmahan2017communication}, FedDyn \cite{acar2021federated}, MOON \cite{li2021model}, FedImpro~\cite{tang2024fedimpro} and FedMut \cite{hu2024fedmut}.
In the \textit{Personalised FL} (pFL) category, which enables centres to develop customised models, we choose FedDM~\cite{xiong2023feddm}, FedProto \cite{tan2022fedproto}, FedHE~\cite{chan2021fedhe}, DESA \cite{huang2024overcoming}, FedD3~\cite{song2023federated}, FedDGM ~\cite{jia2024unlocking} and SGPT \cite{jia2025unlocking}. Additionally, we include a setup where each client trains solely on its own local data without FL (Local). 

\noindent \textbf{Homogeneous Local Model} This setup assumes that each medical centre has a similar computational capacity and is willing to adopt a unified model for their local tasks. Results are shown in Table. \ref{tab:standard} and Table. \ref{tab:pfl}, where we use one of the most commonly used MIL classifiers in the field, CLAM \cite{lu2021data}, as the global model. All FL methods outperform Local, highlighting that inter-centre collaboration can enhance histopathological diagnosis. Compared to standard FL methods, our FedWSIDD achieves the highest global average accuracy. 
Compared to personalised FLs, FedWSIDD shows superior performance, outperforming the previous works that adopt dataset distillation for federated learning \cite{xiong2023feddm,huang2024overcoming,song2023federated,jia2024unlocking}. This demonstrates the effectiveness of FedWSIDD for federated WSI classification.

\noindent \textbf{Heterogeneous Local Model} This setup assumes that different medical centres have varying computational resources, leading them to deploy different MIL classifiers. To simulate this scenario, we define a MIL model pool comprising CLAM \cite{lu2021data}, TransMIL \cite{shao2021transmil}, and ABMIL \cite{ilse2018attention}. Additionally, since class embedding dimensions vary across different MIL classifiers, FedProto cannot be applied. Instead, we use FedHE~\cite{chan2021fedhe}, which shares class logits, for comparison.
Each centre is randomly assigned a classifier from this pool. As shown in Table~\ref{tab:heter}, FedWSIDD not only outperforms pFL methods but also demonstrates greater stability with a lower standard deviation. Furthermore, the p-values for all comparisons are below 0.05, confirming that FedWSIDD’s improvements are statistically significant.

\begin{figure*}[!t]
\centering
\centerline{\includegraphics[width=0.9\textwidth]{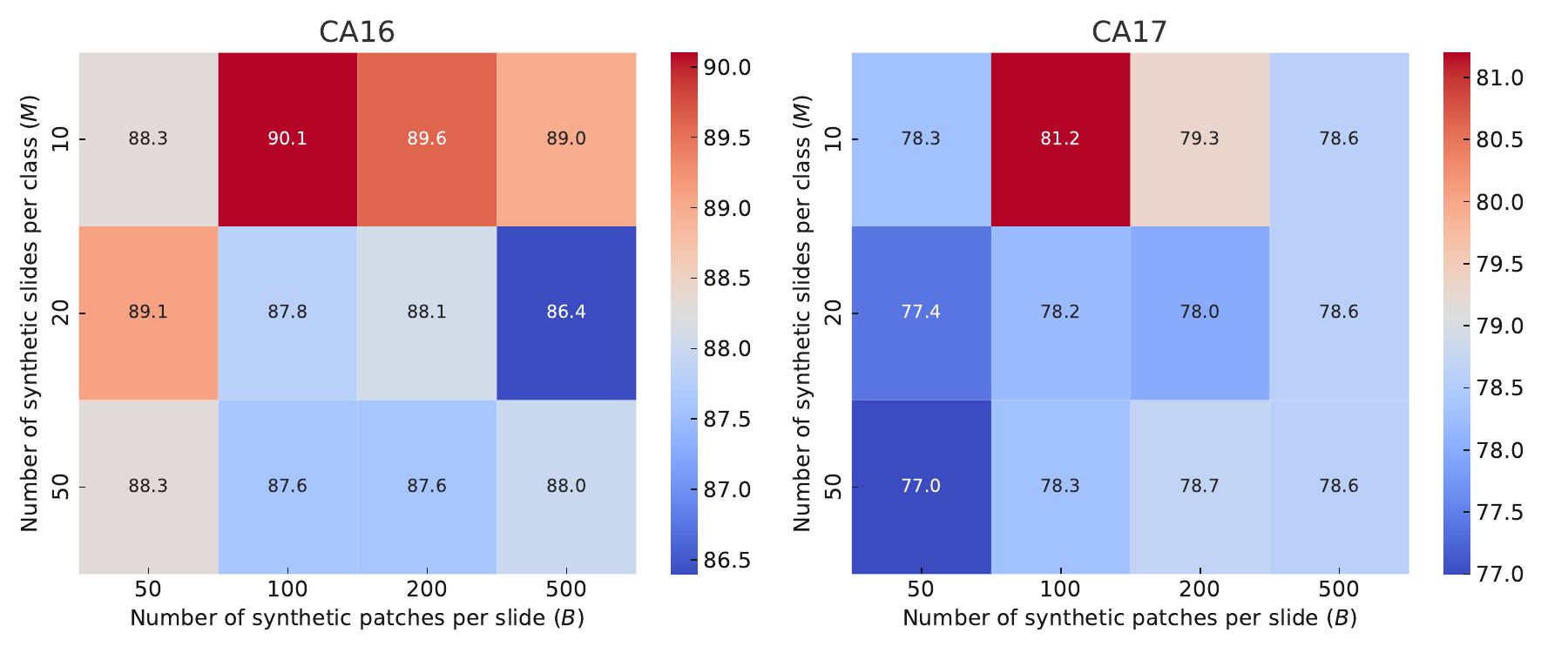}}
\caption{Global averaged accuracy (\%) across centres in CAMELYON16 (CA16) and CAMELYON17 (CA17) with different $M$ and $B$.}
\label{fig:ab}
\end{figure*}
\begin{figure*}[!t]
\centering
\centerline{\includegraphics[width=1.0\textwidth]{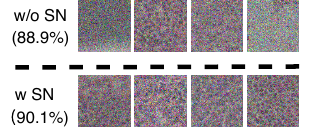}}
\caption{Performance comparison with and without using stain normalisation during dataset distillation.}
\label{fig:sn}
\end{figure*}
\subsection{Ablation studies}
\noindent \textbf{Evaluation of synthetic slide dimension} 
The dimension of the synthetic slide is a crucial hyperparameter, as it not only determines the number of parameters transmitted across clients but also affects the learning complexity of FedWSIDD. We present the performance of FedWSIDD on two datasets under different combinations of the number of synthetic slides per class (\(M\)) and the number of synthetic patches per slide (\(B\)) in Fig.~\ref{fig:ab}. The results suggest that blindly increasing \(M\) and \(B\) does not necessarily lead to improved performance. We attribute this to the limited data variation in WSI datasets, where a smaller set of synthetic samples may suffice to capture the discriminative features of the entire dataset. Furthermore, increasing \(M\) and \(B\) imposes higher computational demands and may hinder the convergence of FedWSIDD.

\noindent \textbf{Effectiveness of stain normalisation}
We compare the global average accuracy across centres on CAMELYON16 with and without stain normalisation in Fig.~\ref{fig:sn}. The results demonstrate that stain normalisation consistently improves the quality of synthetic samples, leading to enhanced model performance. This underscores its crucial role in making dataset distillation algorithms more adaptable to whole-slide image (WSI) datasets.  
Moreover, since the distilled samples are synthetic images, they offer the advantage of enhancing privacy by desensitising sensitive information. However, this synthetic nature may also reduce their explainability to some extent.

\section{Conclusion}
In conclusion, FedWSIDD offers a novel personalised federated learning framework for WSI classification, effectively addressing the challenges of computational constraints in medical centres. By exchanging synthetic slides, FedWSIDD achieves state-of-the-art performance with rapid convergence. Extensive results across multiple WSI datasets show that our approach not only enhances accuracy but also adapts well to both homogeneous and heterogeneous model settings, making it a promising solution for federated learning in resource-limited digital pathology applications.

\bibliographystyle{splncs04}
\bibliography{ref}
\end{document}